# Programmable Biomolecule-Mediated Processors

Jian-Jun SHU[a]*, Zi Hian TAN[a], Qi-Wen WANG[a], and Kian-Yan YONG[a]

[a]School of Mechanical & Aerospace Engineering, Nanyang Technological University, 50 Nanyang Avenue, Singapore 639798

*Email: mjjshu@ntu.edu.sg

**ABSTRACT:** Programmable biomolecule-mediated computing is a new computing paradigm as compared to contemporary electronic computing. It employs nucleic acids and analogous biomolecular structures as information-storing and -processing substrates to tackle computational problems. It is of great significance to investigate the various issues of programmable biomolecule-mediated processors that are capable of automatically processing, storing, and displaying information. This Perspective provides several conceptual designs of programmable biomolecule-mediated processors and provides some insights into potential future research directions for programmable biomolecule-mediated processors.

## ■ INTRODUCTION

Whenever the word "computer" is mentioned, our intuition automatically associates it with an image of a monitor and keyboard, or various technical terms such as central processing unit (CPU), random access memory (RAM), and read-only memory (ROM). This is because we have grown accustomed to the concept of emulating computation through the use of devices commonly referred to as digital computers, which comprise an array of functional electronic components assembled on a silicon substrate. Since the introduction of the first digital computer in the early 1970s, improving its computational ability—processing speed, parallelism, minimization, and energy efficiency—has been the issue of most concern. To meet the ever-increasing demand for processing speed and parallelism, the size of individual transistor elements must be reduced. Doing so allows additional processing units to be packaged on the same silicon die; however, increasing the packaging density always brings problems, including increased power consumption and problematic heat dissipation issues. Moreover, the employment of silicon substrates as basic materials in the manufacture of digital computers always has a negative impact on health and the environment.[1] Most importantly, the entire semiconductor industry is rapidly approaching the physical constraints predicted by Moore's law.[2] Moreover, current computer technology based on silicon materials and binary algorithms is restricted by further component miniaturization[3] and operational speed.[4]

In reality, any device endowed with the three essential functions, namely, information processing, storing, and displaying, can be regarded as a computer.[5] This ratiocination enables scholars from various disciplines to explore other potential alternatives to contemporary electronic digital computers. Among the various intriguing approaches, programmable biomolecule-mediated computing is a feasible method because of biomolecules' appealing features as perfect nanomaterials, including their minuscule size, short structural repetition, and reasonable stiffness.[6] As information storage media, the basic building blocks—deoxyribonucleic acid (DNA), ribonucleic acid (RNA), or protein—are extremely small in size, capable of encoding a single bit of information in a solution of about one cubic nanometer.[7] Furthermore, reasonable stiffness[6] and a quantitative understanding of biomolecular thermodynamics enhance the reliability of DNA as a data storage device.[8] A biomolecule-mediated processor, as an information-processing unit, is capable of executing approximately ten trillion calculations at a time.[9] The reason for this efficiency comes from the fact that, unlike conventional silicon-based digital computers, biomolecule-mediated computing executes calculations in parallel.[10] Most silicon-based digital computers work in a linear fashion, executing one task at a time and repeating similar operations. In contrast, biomolecule-mediated computation executes multiple types of operations simultaneously and stochastically. It is this adaptive intelligence and sophisticated parallel computing that enable synthetic biomolecule-mediated processors to effectively solve complex problems. By comparison, a contemporary silicon-based digital computer could take hundreds of years to achieve the same solution. Furthermore, biomolecule-mediated computing is more energy-efficient as compared to modern computers.

The interdisciplinary research on biomolecule-mediated computing is closely related to the progress in biomolecular engineering. With a better quantitative understanding of biomolecular thermodynamics, additional sophisticated biomolecular structures can be created to increase information storage stability as well as processing speed. In addition, improvements in the quality and rapidly falling costs of synthesizing nucleic acids, coupled with the development of functional enzymes and available laboratory techniques, have provided an additional impetus to the discipline. Nowadays, biomolecule-mediated computing is gradually shifting from the development of limited function biomolecular devices toward the creation of conceptual models of general-purpose biomolecule-mediated logic gates, analogous to the evolutionary history of electronic digital computers. At the same time, operations are no longer restricted to *in vitro* manipulations. On the contrary, the current findings indicate that these experiments can be successfully implemented under sophisticated cellular conditions. In recent years, the entire field has grown rapidly, building an enormous technical barrier for upcoming scholars with different knowledge backgrounds. Therefore, this Perspective attempts to cover the development from early-stage, limited-function biomolecular models to general-purpose, biomolecule-mediated logic circuits. Moreover, the focus of this paper is to clarify the underlying logic behind each stage of development and gradually outline the challenges of interdisciplinary research.

**Silicon-Based Computing.** Nowadays, computers are immensely powerful and can execute millions of calculations per second. They are small in size and affordable for a great many people. It would be pretty astonishing to track the rate at which computers have evolved since the first generation of computers were manufactured around 1941. They were driven by electromechanical components, with instructions provided via punch cards. The second generation of computers were built between 1941 and 1950 through the use of vacuum tubes and capacitors. Vacuum tubes were used as switching elements that defined the state of a computer program. Capacitors enabled the computer to have a memory compartment where intermediate results were stored and fed back to the computing system. As a result, computers shrunk in size from what once occupied an entire room to the space of a large desk.

In the 1950s, vacuum tubes were gradually replaced by transistors, giving way to the third generation of computers. Transistors had many significant advantages over vacuum tubes in computing because they were faster, smaller, less expensive, and more dependable. Transistors, along with other electronic components, were connected together on a semiconductor material, known as an integrated circuit (IC). The computer system on the IC that executed the program was known as the CPU. Previously, each CPU could host only one or a few functions. This meant that people had to manually switch between different ICs to adopt distinct functions. This was an inefficient way of computing. The problem was solved when ICs were manufactured that integrated most or all of their functions. This is the well-known microprocessor, which is now the core of modern fourth-generation computers. Since then, computers have become faster and more compact through the use of tiny transistors with advanced nanotechnology. However, according to Moore's law, there is a limit to the small size of a transistor, as it approaches the size of a single atom.[3]

**DNA-Mediated Computing.** DNA-mediated computing can be very compact because DNA strands are exceedingly small (one bit per cubic nanometer compared to one bit per $10^{12}$ cubic nanometers for modern computers), giving it exciting potential. Computations are also amazingly fast due to parallel processing ($10^{14}$ to $10^{20}$ operations per second compared to $10^8$ to $10^{12}$ operations per second in modern computers). DNA-mediated computing is more energy-efficient than modern computers. An operation typified by a reaction between two DNA-strands uses $5\times10^{-20}$ Joules of energy, compared to $10^{-9}$ Joules in a silicon-based computer. It is worth mentioning that the study of DNA-mediated computing may also lead to a better understanding of a more complex computer—the human brain.[11–13]

DNA-mediated computing has several advantages over silicon-based computing. First, if the ligation of two molecules is considered as a single operation, then $10^{20}$ or more operations per second can be executed. In contrast, the fastest supercomputers can execute approximately $10^{12}$ operations per second. Second, according to the Gibbs free energy,[14] one Joule is sufficient for approximately $2\times10^{19}$ ligation operations. For existing supercomputers, $10^9$ operations per Joule are executed. Last, the information density of DNA is one bit per cubic nanometer. For videotapes, the density of information is approximately one bit per $10^{12}$ nanometers. It is worth mentioning that DNA data storage, as the simplest form of DNA-mediated processor, was listed as one of the top ten emerging technologies in 2019 by the World Economic Forum's annual list;[15] however, the advantage of silicon-based computing lies in the diversity of operations and flexibility in which these operations can be applied. The major advantages and drawbacks of DNA-mediated computing relative to silicon-based computing are compared in Table 1.

**Table 1.** Silicon-based computing versus DNA-mediated computing

| Characteristics | Silicon-based | DNA-mediated |
|---|---|---|
| **Information storage** | one bit per $10^{12}$ cubic nanometers | one bit per cubic nanometer |
| **Processing speed** | $10^8$ to $10^{12}$ operations per second | $10^{14}$ to $10^{20}$ operations per second (ligation) |
| **Energy efficiency** | $10^9$ operations per Joule | $2 \times 10^{19}$ operations per Joule |
| **Computing architecture** | Effective for single operation; multiple cores of CPU for multiple operations at one time (up to six operations) | Ineffective for single operation; naturally effective for massive parallel operations |

In silicon-based digital computing, electronic logic gates are the basic components for analysts to execute computational operations. These electronic gates convert electronic signals into binary codes, which are understandable by silicon-based digital computers. Based on this working principle, silicon-based digital computers can execute diverse types of electronic logic gates to implement various tasks.

Compared with the history of evolving electronic computers, the development of DNA-mediated computers is still at an early stage. DNA-mediated computing is a new computing paradigm that utilizes artificially synthesized nucleic acids and/or analogous biomolecular structures as information-storing and -processing substrates to tackle computational problems. The entire field hinges on two important discoveries—the double helix structure of the DNA molecule in 1953,[16] and the understanding of fundamental biomolecular mechanics in the late 1970s. An integrated symmetrical table for the genetic codes of life created by Shu[17] shows that life formed due to symmetry (template) but evolved due to asymmetry (signature). All of the information encoding life is encapsulated within the tiny nucleus of the cell. This extremely condensed nature of DNA makes it ideal for scaling down computations. The specific Watson-Crick base pairings reflect the unique nature of digital information. DNA-mediated computing can be broadly divided into two categories: limited-function DNA algorithms and logic circuits. For the first approach, computation using sequentially encoded information is demonstrated by solving difficult mathematical problems. This approach follows the traditional dogma that, while DNA stores information about life, the digital information is simply encoded in Watson-Crick base pairings. Afterward, calculations are performed using DNA-manipulating enzymes. The second approach leverages Watson-Crick base pairings to define states rather than to store information. The specificity of the Watson-Crick base pairing, which restricts a single-strand DNA (ssDNA) sequence to hybridize only to its reverse complement, can be used to prime a reaction in the presence of a specific DNA input sequence. This reaction acts like a digital logic gate with two input states, "0" or "1", corresponding to a DNA logic gate with a "match" or "mismatch" state with respect to the target input strand sequence, respectively.

# ■ DNA ALGORITHMS

**Ligation-Based System.** The computational use of DNA molecules was demonstrated[7] to solve the Hamiltonian path problem (Figure 1). Although the solution to the problem demonstrated is trivial and could be easily solved by hand, solving a non-deterministic polynomial-time complete (NP-complete) problem *in vitro* with DNA molecules is significant. NP-complete problems grow with the size of the problem, and it is challenging for computers to obtain a solution for large problems. An inherent advantage of utilizing DNA molecules for computing is the parallelism of chemical reactions. In this demonstration, approximately $10^{14}$ DNA molecules were ligated simultaneously, and the ligation of up to $10^{20}$ DNA molecules was possible by increasing the reaction volume. Compared with the supercomputer speed of $10^{12}$ operations per second in 1994, DNA offers 2-8-fold higher computation speed.

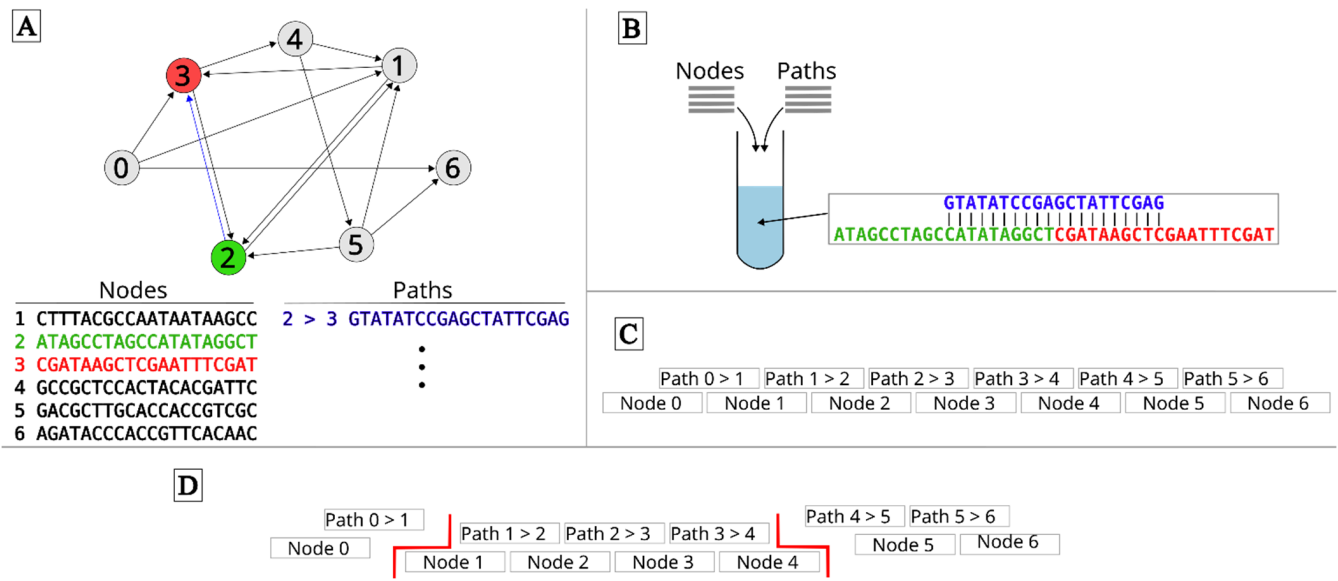

**Figure 1.** Hamiltonian problem solution with the DNA reaction. (A) This Hamiltonian problem has multiple nodes (1-7) and defined directed paths (arrows). The solution is to find a path that starts at node 0 and ends at node 6, while visiting each node only once. Unique DNA sequences of twenty bases are assigned to nodes and defined paths. (B) A sequence of paths must be the unique reverse complement of the two nodes it connects in the Hamiltonian problem. All node and path sequences are added to a polymerase chain reaction mixture for hybridization of path and node DNA molecules. (C) The solution to the path problem is contained in a DNA strand of length 140 bases (seven nodes of twenty bases each). (D) Restriction enzymes can be used to cleave DNA strands at specific sequences to define the start and end points (boundaries) of a path. If the desired path starts at node 1 and ends at node 4, the restriction enzymes can be used to cleave at the sequence of [path 0>1 and node 1] and [node 4 and path 4>5].

The technique has been extended to many other mathematical problems: Boolean satisfiability problem,[10] addition problem,[18] maximal clique problem,[19] Chinese postman problem,[20] maximum matching problem,[21] traveling salesman problem,[22] maximum cut problem,[23] clustering problem,[24] bin-packing problem,[25] and assignment problem.[26] However, all these implementations have limited capabilities, as they are restricted to solving certain classes of combinatorial optimization problems.

**Sequence Editing System.** In addition to ligation (joining), DNA strands can also be cleaved with restriction enzymes. This technique, when used in DNA-mediated computing devices, opens up many more possibilities in terms of computational complexity.[27] In addition to setting the minimum conditions to be met, boundary conditions can also be specified (Figure 1D). Any DNA-strands in solution beyond the boundaries can be simply cleaved and filtered out by employing gel electrophoresis. A programmable and automatic computing machine composed of biomolecules was built around this technique,[28] in which encoded input strands were decoded through a series of loops. Within each loop, a portion of the strand was cleaved if it matched a restriction enzyme recognition site. This process was repeated until the input strand was cleaved to the end or no restriction sites were detected, and the decoded output was read using gel electrophoresis. In addition, the same technique was employed to solve the strategic assignment problem.[26] Restriction enzymes were used to accelerate the process, as demonstrated by a schema-based DNA computing algorithm applicable for graphics processing units.[29] The same principle has been employed in the formation of *in vitro* molecular machine-learning algorithms, opening the way for solving machine-learning problems.[30] Besides restriction enzymes, the use of clustered regularly interspaced short palindromic repeats (CRISPR) allows for more precise cleavage of target sequences.[31]

## ■ DNA LOGIC CIRCUITS

Limited function DNA algorithms, specified in the first category, maximize the advantages of DNA-mediated computing, namely, massively parallel computing capabilities; however, this application is restricted to certain categories of mathematical problems. The second category emphasizes the development of DNA logic circuits, which are assembled from a series of functional logic gates. Analogous to the case of contemporary digital computers, designing appropriate DNA logic circuits is believed to be a viable starting point for the development of general-purpose DNA-mediated computers. Logic circuits are assembled from a series of individual functional units called logic gates, which are capable of executing simple Boolean logic operations. Many other early-stage designs of DNA-mediated logic gates were summarized.[27,32–34] In general, DNA logic circuit designs can be classified into two categories: tiling and DNA strand displacement (DSD).

**Tiling System.** The tiling system was used to emulate an early form of the Turing machine,[35] in which programs were represented on magnetic tape.[36] The Turing machine used symbols to provide a readout based on the order in which the holes were punched in the tape. A separate set of symbols was attained by moving the point where the machine started to read. The starting point was called the controller state, which together with the symbol was referred to as the configuration. Therefore, the configuration was changed by changing the controller state.

DNA self-assembled molecules, known as tiles, were used to represent symbols and controller states. A configuration was a row of tiles. To change the configuration, a new row of tiles was stacked on top of the initial row[37] in a manner determined by the Wang tiles,[38] which were square tiles with colored edges. This arrangement places edges of the same color side by side, creating an aperiodic pattern on a plane. A set of the seven Wang tiles, each with a unique combination of five selected colors, for emulating an exclusive disjunction (XOR) logic is show in Figure 2. Each DNA tile contains a reporter strand that extends diagonally across the tile and ligates to the reporter strand of other connected tiles. To obtain the assembled outputs, a polymerase chain reaction (PCR) is performed with specific primers for each input (X1, X2, X3, X4) and output (Y1, Y2, Y3, Y4).[39] Atomic force microscopy (AFM) can also be used to detect the physical location of tiles with overlapping distinct features.[40] Beyond XOR logic, the tiling system was also used to perform other DNA-mediated logic gates[39,41] and arithmetic calculations. The latter included counting,[42] addition,[40] multiplication,[43] subtraction,[40] and division,[44] as well as elementary functions.[45]

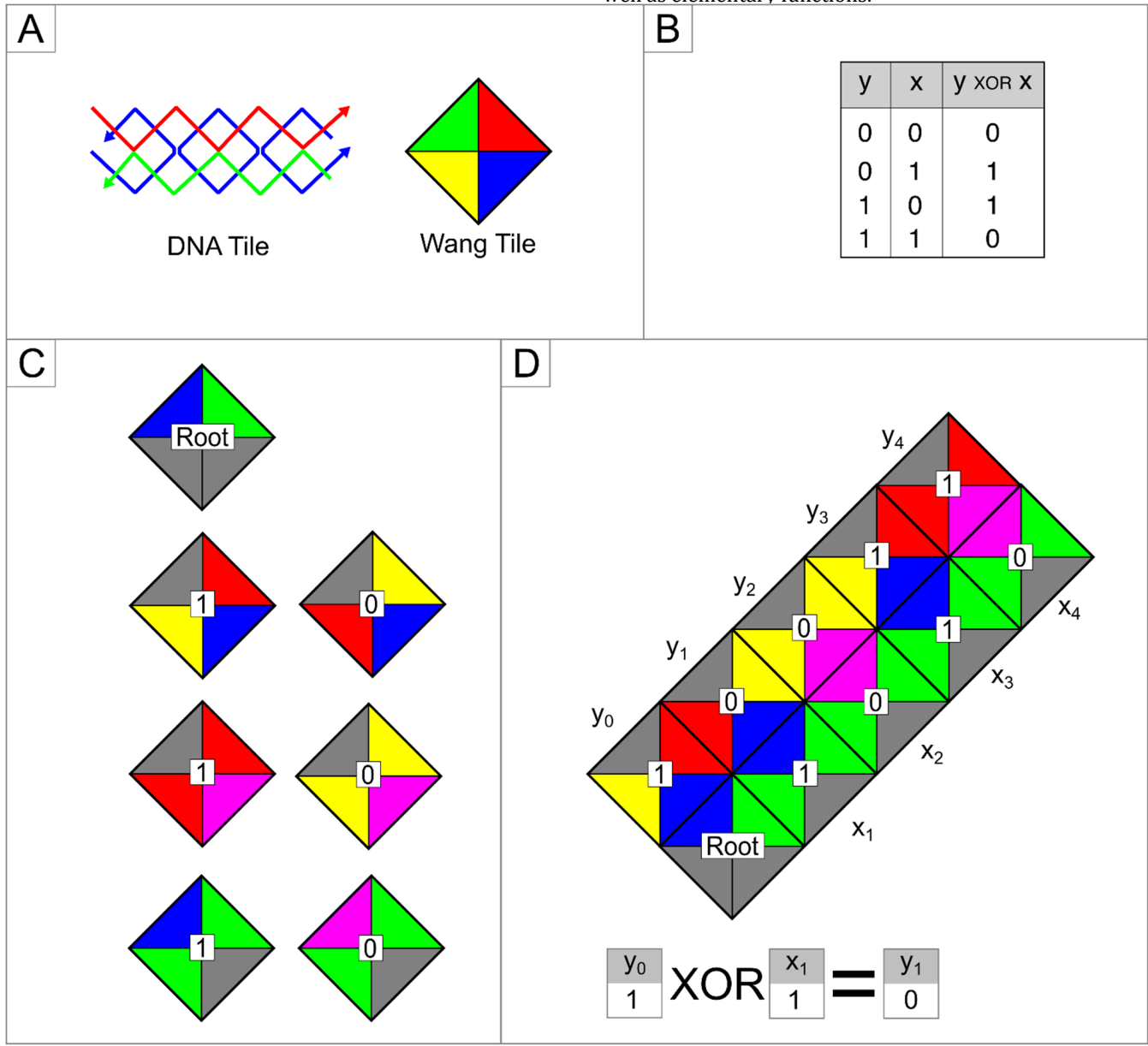

**Figure 2.** (A) DNA assembly with four single-stranded ends for hybridization to other tiles. DNA ends are designed to be specific. Four-sided DNA assembly can be represented as a Wang tile. Each side of the Wang tile is color-coded, and only matching colors fit together, representing the specificity of DNA single-strand end-binding. (B) Truth table for exclusive disjunction (XOR) logic. (C) Seven Wang tiles with specific color combinations are needed to emulate XOR logic. (D) The Wang tile assembly performed an XOR logic computation. X tiles are inputs, and Y tiles are assembled accordingly. The assembly process starts with the bottom left root tile. $X_1$ tile of "1" is introduced. If the first $Y_0$ tile is a '"1", the Wang tile rule dictates that only a Y tile of "0" can be assembled to match the color. Hence, $Y_1$ is "0" according to the logic table. Logic computation is cascading as [$Y_{(i-1)}$ XOR $X_i$=$Y_i$]. The next computation is: $Y_1$ XOR $X_2$=$Y_2$.

Challenges for tiling systems include determining the minimum type of tiles required to generate a solution, how quickly the tiling can be assembled, and whether a solution can be successfully generated for nondeterministic computations.[46] One of the biggest limitations of the tiling system is the writing and reading of information. The DNA strands in a tile are designed to be specific for self-assembly and binding to other tiles. This specificity increases the difficulty of designing strands for multiple tiles and is not suitable for multiple logical operations. After the sequence is designed, the DNA strands must be synthesized for self-assembly. Obtaining results from a tiling system is not immediate, as it requires separate DNA molecular techniques or AFM. These time-consuming constraints are further compounded by the irreversible nature of DNA ligation, limiting the costly synthesized DNA circuits to a single use. Resettable kinetics were demonstrated by applying an excess of ssDNA blocker to outcompete the input of the corresponding binding strand.[47] This addresses the issue of wasted resources and enables logic circuits to be designed once and reused; however, the time and resource costs of the tiling system far outweigh the ability to perform complex computations.

**DNA Strand Displacement System.** Moving from the tiling system, energy-free and enzyme-free reactions are desirable features for DNA computing without a complex design. DSD was demonstrated[48] to occur without additional energy input during natural hybridization and was employed[49] to implement AND, OR, and NOT gates, signal restoration, amplification, feedback, and cascading circuits (Figure 3 top). Compared to the ligation and restriction enzyme method, DSD allows multi-stage computations *i.e.*, the ssDNA output calculated in the first step can be used as the input for the next step. Multilevel computation enables more complex functions to be performed. Other advantages of DSD include sensitivity, simplicity, programmability, and ability to use enzyme-free reactions. Since then, DSD has been applied in various fields,[50–52] such as nanomachines, origami assembly, medical sensing, and diagnostics.[53–55] The physical chemistry and various applications of DSD have been discussed.[52] This section focuses on recent applications of DSD on DNA logic gates.

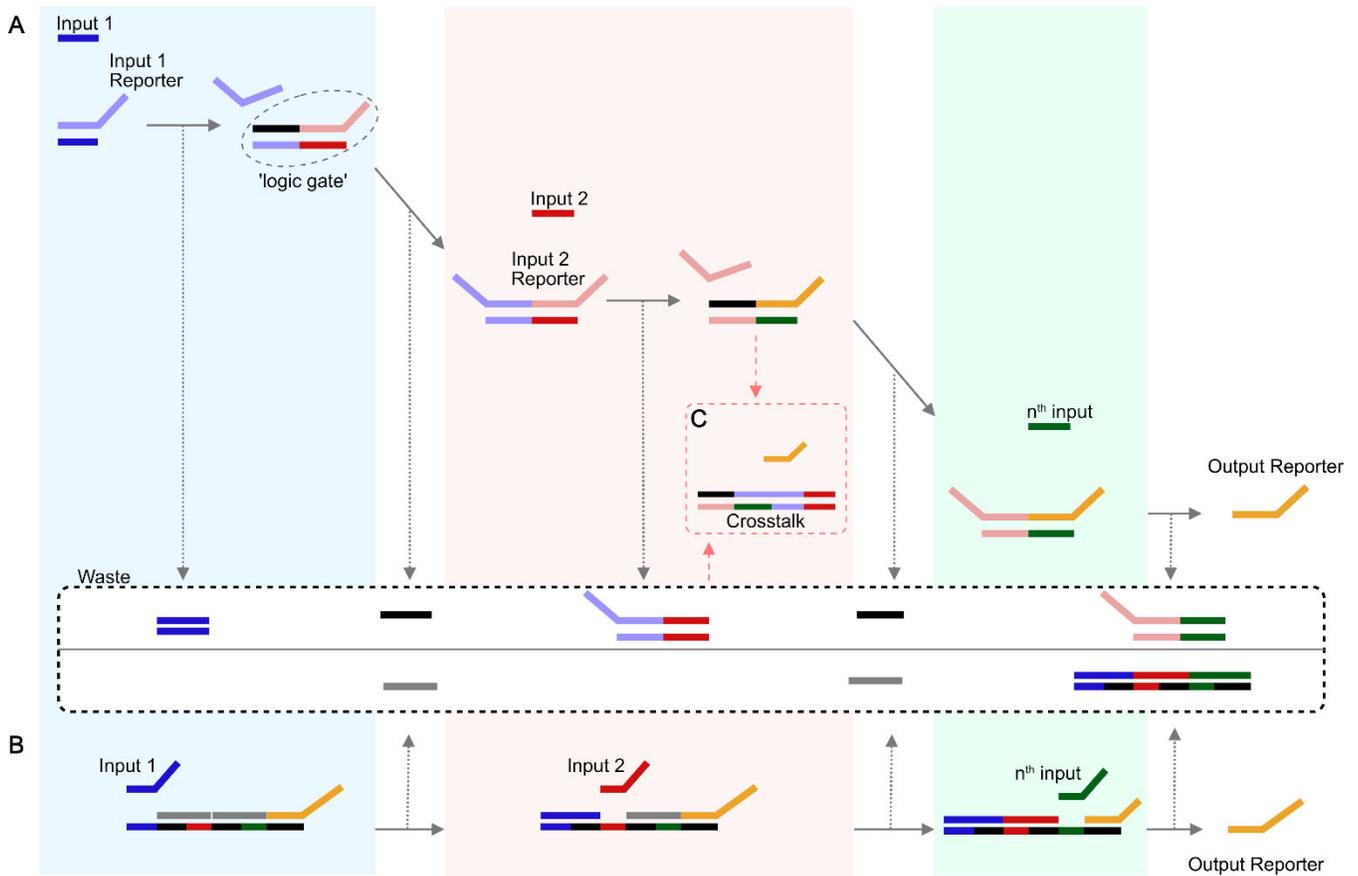

**Figure 3.** (A) Cascading DNA strand displacement (DSD) logic gates: Each single-stranded DNA (ssDNA) input requires a DSD unit that acts as a logic gate, ready to accept the input toehold and release the reporter strand to form the next logic gate for subsequent inputs. The input 2 stage requires the input 1 reporter to signal the completion of the input 1 detection in preparation for the input 2 toehold. By designing more stages to form a cascading repeat circuit, the circuit can accept multiple inputs. (B) Multi-input DSD logic gate: For the same logic operation, only a single assembly of double-stranded DNA with a toehold region corresponding to the ssDNA input is required. Since intermediate reporting strand is eliminated, fewer unique toehold sequences are required, and therefore a simpler toehold design. Multi-input DSD logic gates generate less DNA waste than cascading DSD. (C) The wasted DNA byproducts of DSD can create unwanted pathways, releasing intermediate signals or output reporters. Multi-input DSD produces less waste DNA than cascading DSD.

Logic gates are the building blocks for the functional design of a complex computing device. One of the hurdles in DNA computing is that logic circuits have to be designed *de novo* for an application. Standardized units were designed to mimic gene regulatory networks with feedforward and feedback modules (genelets) for simple kinetics of gene activation, activated by RNA repressors and coactivator strands.[56] To create a modular unit with minimal crosstalk and noise, a hairpin clamp (HPC5) was designed at the input to bind only the RNA repressor and coactivator strands. Although the demonstrated logic is between two states ("ON" and "OFF"), with a simple reaction kinetics output, the potential advantage of a standardized module enables a quick and modular complete logic design; however, when compared to silicon circuits, DSD logic gates have ssDNA waste that must be managed through a constant flow of solutions. In a complex system where designing multiple unique toehold sequences is challenging, discarded DNA strands can interact with the toehold region and release the output strand (leakage), thereby reducing the sensitivity and reliability of the circuit (Figure 3C). Lowering the concentration helps reduce leakage but also slows down the reaction rate. Several efforts were done to reduce leakage by introducing error-correction methods;[57] however, such approaches tend to introduce design complexity that reduces the use of DSD in complex logic applications, limiting DSD logic gates to simple and small networks compared to silicon digital circuits.

Cascading DSD leakage compensation is a counterbalance to the toehold principle, where the added complexity makes the unique toehold sequence design difficult. If multiple inputs are required, a multi-input DSD logic gate (Figure 4) was demonstrated to eliminate the need for cascading logic,[58] performing multi-input AND logic on a single DNA unit. It uses a single DSD "logic gate" unit for multiple inputs rather than cascading multiple DSD units per input (Figure 3). This approach has two advantages: fewer nucleotides are synthesized in the same operation, and fewer waste DNA byproducts are produced, thus reducing leakage. Nonetheless, the fundamental requirement of unique toehold sequences still limits the computational scale of multi-input DSD. Still, one major advantage of DNA computing is parallelism, not computational complexity or depth, speeding up multi-input and massively parallel computing, such as the tic-tac-toe game demonstrated. Massive parallelism makes DNA, and in particular DSD computation, well-suited for emulating the nodes of artificial neural networks (ANNs), where each node has low computational complexity and is not restricted by a toehold (Figure 5). ANN, demonstrated with *in silico*-designed ssDNA,[59] was shown to work with cancer-associated miR-2 and miR-31 microRNAs (miRNAs).[60]

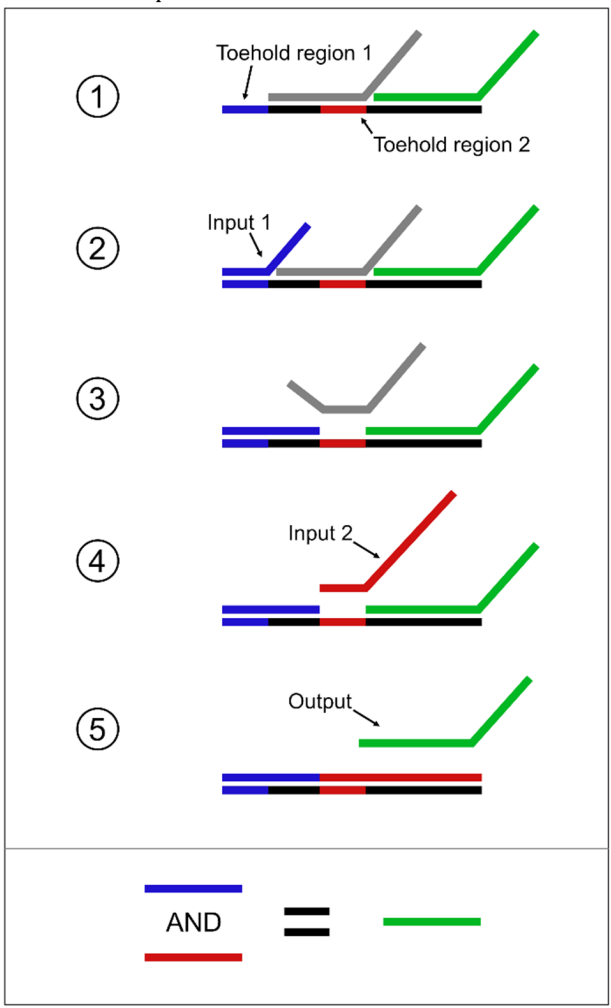

**Figure 4.** DNA strand displacement (DSD) used to emulate conjunction (AND) logic. AND logic outputs a positive signal when both input 1 AND input 2 are present. 1: DNA assembly with two toehold regions for two inputs, respectively. 2: The first input single-strand DNA (ssDNA) emerges and attaches to toehold region 1 (blue). 3: The DSD process occurs and displaces excess ssDNA, revealing toehold region 2 (red). 4: Input 2 ssDNA binds to revealed toehold region 2 and initiates DSD process. 5: After complete displacement by input 2 ssDNA, the final excess ssDNA (green) is the result of input 1 AND input 2 ssDNA. Overall process emulates AND where only in the presence of input 1 and 2 ssDNA is output ssDNA released.

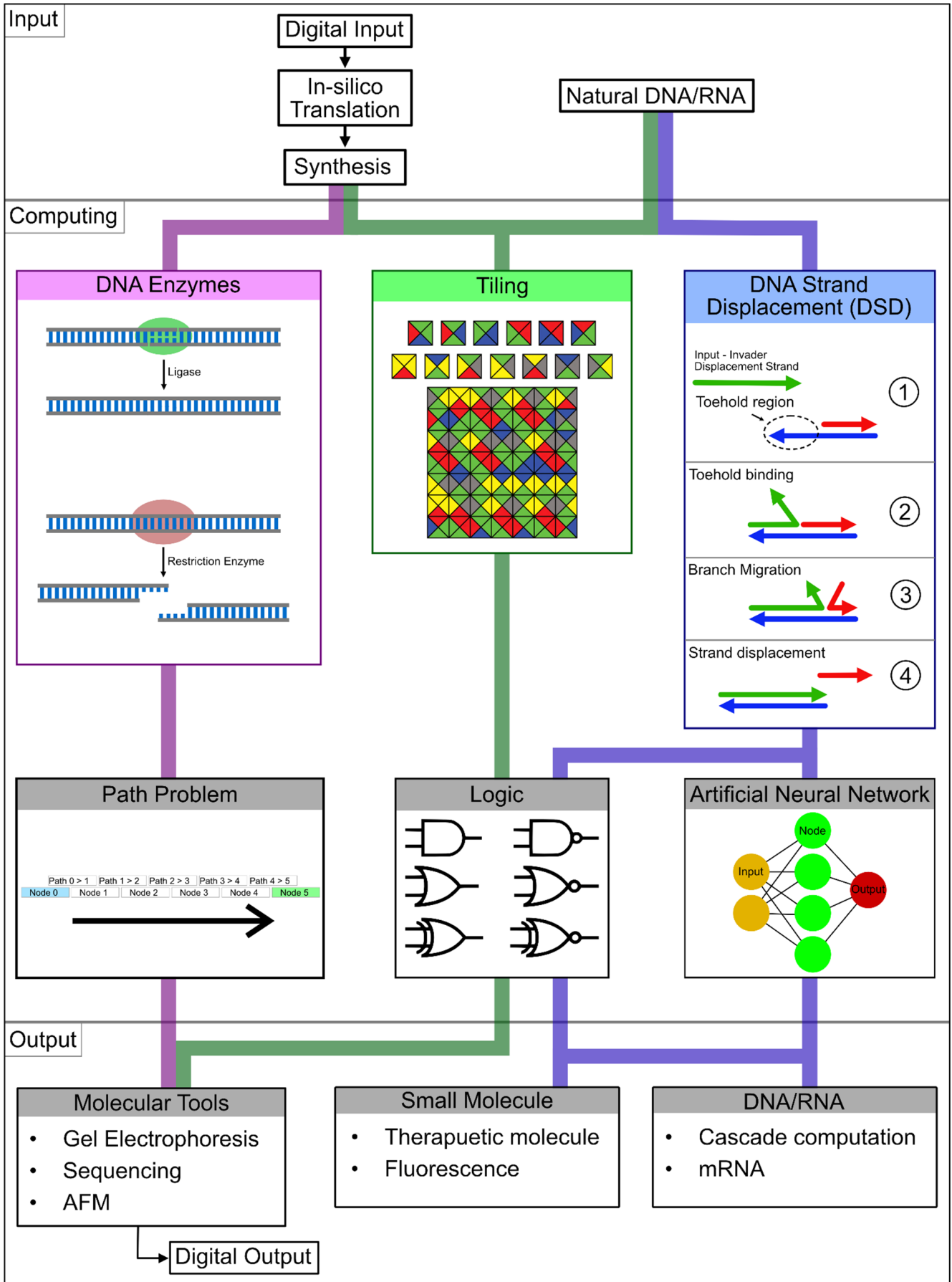

**Figure 5.** Overview of DNA-mediated computing. Each method (deoxyribonuclease, tiling, and DNA strand displacement (DSD)) may be restricted to a limited input, computing (path problem, logic, or artificial neural network (ANN)), and output. Input and output combined form the interface of a method. For DSD, input may be synthetic or natural (cancer miRNAs and viral RNA). The DSD computing task can be directed to logic gates or ANNs and lead to either a small-molecule release or DNA/RNA release.

In addition to parallel computing, DSD can be used in fields that require only simple computations using natural DNA rather than digital data as input. Circulating miRNA is a good biomarker for rapid early detection, prognosis, and prediction of cancer.[61] DSD-based cancer diagnosis was demonstrated with addition, subtraction, and multiplication gates.[55] The result of computation was expressed as fluorescence intensity or small-molecule release.[62] This has some potential applications where the specificity of single or multiple miRNA biomarkers can be detected to release therapeutic molecules *in vivo*. DSD does not require additional proteins and enzymes for computing. Hence, it can be applied *in vivo* without introducing exogenous proteins into the subject, which has potential applications in biosensing.[63,64] Beyond cancer, DSD has also been developed for viral RNA detection, such as *Zika virus*.[65] Combined with DSD to perform isothermal RNA amplification of target viral RNA for detection, the entire process can be performed on paper, reducing the time, cost, and expertise of nucleic acid detection.

## ■ CONCLUSION

In recent years, the research focus of DNA-mediated computing has gradually been shifting from the development of limited-function DNA algorithms to the demonstration of practical DNA-mediated logic gates. Because the specificity of the Watson-Crick base pairs mirrors the distinct nature of digital information, the use of DNA for computation is attractive. DNA computing can be achieved through two main approaches: computing using sequential information encoding and computing using sequential encoding states. Both approaches have natural limitations.

Hence, any processing or computation is initially done by synthesizing the designed sequence (input) followed by specific DNA enzymes such as ligase, restriction enzyme, or CRISPR (computation). As the information is stored in the sequence itself, gel electrophoresis or sequencing is required to extract the results. Although this approach has the highest storage density and parallel processing potential, the net speed of computation is limited by the synthesis and sequencing throughput and cost. Computing with enzymes also has the drawback that enzymes can affect subsequent logical operators by becoming a "noise signal". So far, there is no workable way to perfectly remove this "noise signal" from the solution.

As such, the DNA sequence does not store information but is used to match a specific input. The entire toehold region encodes only two states; thus, it is not as information-dense as the first approach. Nevertheless, it avoids the introduction of enzymes ("noise signals") because its input signals, logical operators, and output signals are all nucleic acids. Therefore, such logic circuits are always referred to as enzyme-free logic circuits. Unfortunately, the potential of DSD-mediated logic circuits for complex computation is extremely limited. This is because the process of toehold DSD is slow. In addition, a three-level logic circuit consisting of eight logic gates contains 130 different oligonucleotides. It is hard to imagine that there might be sufficient nucleic acids on this planet to construct a general-purpose programmable DNA-mediated processor. Additionally, the DSD process generates waste DNA, which must be managed during the computational process. While the two approaches can be different in working principle, both share the same limitations. Some of the limitations of biomolecule-mediated computing devices together with challenges to the discipline are as follows:

**Large-Scale Limitations.** DNA computing theoretically offers massively parallel computing capabilities with high information density; however, it is unlikely to reach experimental limitations, because the number of nucleic acid templates required to form a complete database grows exponentially with problem size. The example of the traveling salesman problem is used to demonstrate the experimental limitations of algorithms of this nature, with respect to the number of vertices. Suppose the traveling salesman problem contains sixty-two cities. The goal is to determine the shortest path through which a salesperson can pass through all cities exactly once. There are sixty-two cities, and the number of paths is $62! \approx 3.15 \times 10^{85}$. Even with the latest supercomputers, it takes a long time to render the distance of all the paths and determine the solution to the problem (*i.e.*, shortest path).

On the other hand, based on the proposed biomolecule-mediated computing device,[22] to activate the signal, 2015 unique DNA-strands must be designed and synthesized; however, designing a large number of long DNA-strands, with the same melting temperature and avoiding undesired mismatches, is always a challenge. Another, more critical issue is that it is hardly feasible to render a solution to this problem due to the substantial number of DNA-strands that need to be synthesized first. Mathematically speaking, to have 2015 unique DNA-strands would require each strand to have a minimum length of 142 bp. Assuming that each vertex (city) and potential path of the traveling salesman problem is represented by a DNA stand with a length of 150 bp, a complete path, passing through all cities exactly once, might be 18450 bp in length. The length of the resultant DNA-strand is close to the upper limit of agarose gel electrophoresis, which simply means that it may take an excessive amount of time to extract it from the mixture. In addition, approximately $6.37 \times 10^{67}$ kg of DNA is required to ensure that the solution pool contains at least 100 copies of the DNA strands encoding the solution path.

**Irreversibility.** In the majority of the proposed biomolecule-mediated computing devices, the DNA strands encoding the solution to the problem can be collected at the end of the process; however, the obtained DNA strands cannot be reused for computational purposes. It specifies that the information processing conducted by the proposed conceptual models is irreversible. At this stage, although the DNA template cannot be recycled, the amount of DNA synthesized (after signal activation) can be manually duplicated by using PCR prior to the formation of the database. By doing this, the cost of running experiments can be confined to a reasonable level. As a trade-off, additional time must be invested in the duplication process, and the duration can be really extensive as the scale of the problem increases. The development of reconfigurable computing units[47] is necessary for practical computing devices, and even more so for DNA-mediated computers, because synthesis and sequencing are costly and time-consuming to produce.

**Human and Experimental Errors.** Biomolecule-mediated computing devices are still in the initial stages of development; thus, a majority of the processes implemented by the individual functional units are not yet fully automated. Therefore, the determination of the correct answer to the selected problem is highly dependent on manual operations. Manual intervention is likely to always lead to unavoidable human and experimental errors, including problematic nucleic acid template design, improper laboratory setup (solution temperature and time for each process), and solution contamination, which can lead to highly error-prone biomolecule-mediated computing devices in practice.

Problematic nucleic acid template design means that design motifs utilizing a high proportion of DNA sequences are remarkably similar to each other. This is likely to cause mismatches in the non-complementary DNA strands and lead to incorrect answers to questions. Such problems can be minimized by deliberately selecting the sequence and length of the DNA, as long as a set of criteria is followed.[66,67] On the other hand, the implementation of the design protocol may significantly limit the number of DNA templates available, thus limiting the scale of the problem that the proposed device is able to tackle. Alternatively, to prevent mismatches of non-complementary DNA strands, a feedback system was employed.[68] Inappropriate laboratory setup and solution contamination can be avoided by fully automating the entire process, from database formation to signal display.

**Interface.** As with any computing device, the interface plays a significant role in its usability. For DNA to perform any computation task, the problem is transformed from the digital domain into sequences (for the first approach). Although there are software packages that can design questions and convert them into sequences, the biggest limitation, in terms of time and cost, remains the synthesis of DNA. Using photonics such as fluorescence can eliminate the need to sequence simple output computations.[69] For multi-dimensional output, the introduction of surface-based techniques into DNA-mediated computing provides additional flexibility in the design of output representations.[70,71] Interestingly, for complex computations like the path problem, the interface is limited to digital computers, which are necessary for everything from *in silico* design and translation of mathematical problems to sequencing to reading results.

However, this limitation is, in turn, an advantage for computing in biological environments, allowing direct biological interfaces. The ability to receive RNA as input has potential in diagnostics and biosensing. Cancer miRNAs can be detected without the use of PCR,[55] opening up a new field of testing methods. More crucially, the direct biological interface allows *in vivo* computation to be combined with the direct release of therapeutic molecules[62] or RNA signals *in vivo*. Simple *in vivo* DSD logic can be applied to high-risk patients for *in vivo* sensing of diagnostic or predictive cancer biomarkers[72] and trigger-reporting signals or initial stage small-molecule therapy. Higher sensitivity can only be achieved when multiple biomarkers are present in an AND logic and within predefined threshold limits.[59] Several studies have demonstrated the feasibility of cellular computation.[73,74] The future of biomolecule-mediated computing may not be a replacement for silicon, but the use of the inherent advantages of programmable biomolecule-mediated computing to interface with native DNA or *in vivo* is ideally suited to unlock new medical technologies.

## ■ AUTHOR INFORMATION

### Corresponding Author

* **Jian-Jun SHU** – School of Mechanical & Aerospace Engineering, Nanyang Technological University, 50 Nanyang Avenue, Singapore 639798; orcid.org/0000-0003-1557-4077; Email: mjjshu@ntu.edu.sg

### Authors

**Zi Hian TAN** – School of Mechanical & Aerospace Engineering, Nanyang Technological University, 50 Nanyang Avenue, Singapore 639798

**Qi-Wen WANG** – School of Mechanical & Aerospace Engineering, Nanyang Technological University, 50 Nanyang Avenue, Singapore 639798

**Kian-Yan YONG** – School of Mechanical & Aerospace Engineering, Nanyang Technological University, 50 Nanyang Avenue, Singapore 639798

### Notes

The authors declare no competing financial interest.

## ■ ACKNOWLEDGMENTS

This research is supported by the Ministry of Education, Singapore, under its Academic Research Fund Tier 1 (RG75/20).

## ■ REFERENCES

(1) Chepesiuk, R. Where the chips fall: Environmental health in the semiconductor industry, *Environmental Health Perspectives* **1999**, *107*(*9*), A452−A457.
(2) Moore, G. E. Cramming more components onto integrated circuits, *Electronics* **1965**, *38*, 114−117.
(3) Fuechsle, M.; Miwa, J. A.; Mahapatra, S.; Ryu, H.; Lee, S.; Warschkow, O.; Hollenberg, L. C. L.; Klimeck, G.; Simmons, M. Y. A single-atom transistor, *Nature Nanotechnology* **2012**, *7*(*4*), 242−246.
(4) Katz, E. Biocomputing−Tools, aims, perspectives, *Current Opinion in Biotechnology* **2015**, *34*, 202−208.
(5) Danchin, A. Bacteria as computers making computers, *FEMS Microbiology Reviews* **2009**, *33*(*1*), 3−26.
(6) Seeman, N. C. DNA in a material world, *Nature* **2003**, *421*(*6921*), 427−431.
(7) Adleman, L. M. Molecular computation of solutions to combinatorial problems, *Science* **1994**, *266*(*5187*), 1021−1024.
(8) Church, G. M.; Gao, Y.; Kosuri, S. Next-generation digital information storage in DNA, *Science* **2012**, *337*(*6102*), 1628−1628.
(9) Dong, Y. F.; Dong, C.; Wan, F.; Yang, J.; Zhang, C. Development of DNA computing and information processing based on DNA-strand displacement, *Science China-Chemistry* **2015**, *58*(*10*), 1515−1523.
(10) Lipton, R. J. DNA solution of hard computational problems, *Science* **1995**, *268*(*5210*), 542−545.
(11) Condon, A. DNA and the brain, *Nature* **2011**, *475*(*7356*), 304−305.
(12) Shu, J.-J.; Wang, Q.-W.; Yong, K.-Y.; Shao, F.; Lee, K. J. Programmable DNA-mediated multitasking processor, *Journal of Physical Chemistry B* **2015**, *119*(*17*), 5639−5644.
(13) Wong, J.R.; Lee, K.J.; Shu, J.-J.; Shao, F. Magnetic fields facilitate DNA-mediated charge transport, *Biochemistry* **2015**, *54*(*21*), 3392−3399.
(14) Gibbs, J. W. A method of geometrical representation of the thermodynamic properties of substances by means of surfaces, *Transactions of the Connecticut Academy of Arts and Sciences* **1873**, *2*(*2*), 382−404.
(15) DiChristina, M.; Meyerson, B. S. Top 10 emerging technologies 2019, *Scientific American* **2019**, *321*(*6*), 27−28.
(16) Watson, J. D.; Crick, F. H. C. Molecular structure of nucleic acids−A structure for deoxyribose nucleic acid, *Nature* **1953**, *171*(*4356*), 737−738.
(17) Shu, J.-J. A new integrated symmetrical table for genetic codes, *BioSystems* **2017**, *151*, 21−26.
(18) Guarnieri, F.; Fliss, M.; Bancroft, C. Making DNA add, *Science* **1996**, *273*(*5272*), 220−223.
(19) Ouyang, Q.; Kaplan, P. D.; Liu, S. M.; Libchaber, A. DNA solution of the maximal clique problem, *Science* **1997**, *278*(*5337*), 446−449.
(20) Yin, Z. X.; Zhang, F. Y.; Xu, J. A Chinese postman problem based on DNA computing, *Journal of Chemical Information and Computer Sciences* **2002**, *42*(*2*), 222−224.
(21) Wang, S. Y. DNA computing of bipartite graphs for maximum matching, *Journal of Mathematical Chemistry* **2002**, *31*(*3*), 271−279.
(22) Lee, J. Y.; Shin, S.-Y.; Park, T.H.; Zhang, B.-T. Solving traveling salesman problems with DNA molecules encoding numerical values, *BioSystems* **2004**, *78*(*1-3*), 39−47.
(23) Xiao, D. M.; Li, W. X.; Zhang, Z. Z.; He, L. Solving maximum cut problems in the Adleman−Lipton model, *BioSystems* **2005**, *82*(*3*), 203−207.
(24) Bakar, R. B. A.; Watada, J.; Pedrycz, W. DNA approach to solve clustering problem based on a mutual order, *BioSystems* **2008**, *91*(*1*), 1−12.
(25) Sanches, C. A. A.; Soma, N. Y. A polynomial-time DNA computing solution for the bin-packing problem, *Applied Mathematics and Computation* **2009**, *215*(*6*), 2055−2062.
(26) Shu, J.-J.; Wang, Q.-W.; Yong K.-Y. DNA-based computing of strategic assignment problems, *Physical Review Letters* **2011**, *106*(*18*), 188702.
(27) Miyamoto, T.; Razavi, S.; DeRose, R.; Inoue, T. Synthesizing biomolecule-based Boolean logic gates, *ACS Synthetic Biology* **2013**, *2*(*2*), 72−82.
(28) Benenson, Y.; Paz-Elizur, T.; Adar, R.; Keinan, E.; Livneh, Z.; Shapiro E. Programmable and autonomous computing machine made of biomolecules, *Nature* **2001**, *414*(*6862*), 430−434.
(29) Hwang, K.-S.; Chen, Y.-J.; Jiang, W.-C. A schema-based DNA algorithm applicable on graphics processing units, *IEEE Access* **2016**, *4*, 2498−2506.
(30) Lee, J.-H.; Lee, S.H.; Baek, C.; Chun, H.; Ryu, J.-H.; Kim, J.-W.; Deaton, R.; Zhang, B.-T. *In vitro* molecular machine learning algorithm *via* symmetric internal loops of DNA, *BioSystems* **2017**, *158*, 1−9.
(31) Zhang, J. Y.; Liu, C. C. CRISPR-powered DNA computing and digital display, *ACS Synthetic Biology* **2021**, *10*(*11*), 3148−3157.
(32) Stojanovic, M. N.; Mitchell, T. E.; Stefanovic, D. Deoxyribozyme-based logic gates, *Journal of the American Chemical Society* **2002**, *124*(*14*), 3555−3561.
(33) Balzani, V.; Credi, A.; Venturi, M. Molecular logic circuits, *ChemPhysChem* **2003**, *4*(*1*), 49−59.
(34) Hockenberry, A. J.; Jewett, M. C. Synthetic *in vitro* circuits, *Current Opinion in Chemical Biology* **2012**, *16*(*3-4*), 253−259.
(35) Turing, A. M. On computable numbers, with an application to the Entscheidungsproblem, *Proceedings of the London Mathematical Society* **1937**, *42*, 230−265.
(36) Fujibayashi, K.; Hariadi, R.; Park, S. H.; Winfree, E.; Murata, S. Toward reliable algorithmic self-assembly of DNA tiles: A fixed-width cellular automaton pattern, *Nano Letters* **2008**, *8*(*7*), 1791−1797.
(37) Winfree, E.; Liu, F. R.; Wenzler, L. A.; Seeman, N. C. Design and self-assembly of two-dimensional DNA crystals, *Nature* **1998**, *394*(*6693*), 539−544.
(38) Wang, H. Proving theorems by pattern recognition II, *Bell System Technical Journal* **1961**, *40*(*1*), 1−41.
(39) Mao, C. D.; LaBean, T. H.; Reif, J. H.; Seeman, N. C. Logical computation using algorithmic self-assembly of DNA triple-crossover molecules, *Nature* **2000**, *407*(*6803*), 493−496.
(40) Tandon, A.; Song, Y.; Mitta, S. B.; Yoo, S.; Park, S.; Lee, S.; Raza, M. T.; Ha, T. H.; Park, S. H. Demonstration of arithmetic calculations by DNA tile-based algorithmic self-assembly, *ACS Nano* **2020**, *14*(*5*), 5260−5267.
(41) Carbone, A.; Seeman, N. C. Circuits and programmable self-assembling DNA structures, *Proceedings of the National Academy of Sciences of the United States of America* **2002**, *99*(*20*), 12577−12582.
(42) Barish, R. D.; Rothemund, P. W. K.; Winfree, E. Two computational primitives for algorithmic self-assembly: Copying and counting, *Nano Letters* **2005**, *5*(*12*), 2586−2592.
(43) Brun, Y. Arithmetic computation in the tile assembly model: Addition and multiplication, *Theoretical Computer Science* **2007**, *378*(*1*), 17−31.
(44) Zhang, X. C.; Wang, Y. F.; Chen, Z. H.; Xu, J.; Cui, G. Z. Arithmetic computation using self-assembly of DNA tiles: Subtraction and division, *Progress in Natural Science-Materials International* **2009**, *19*(*3*), 377−388.
(45) Raza, M. T.; Tandon, A.; Park, S.; Lee, S.; Nguyen, T. B. N.; Vu, T. H. N.; Jo, S.; Nam, Y.; Jeon, S.; Jeong, J.-H.; Park, S. H. Demonstration of elementary functions *via* DNA algorithmic self-assembly, *Nanoscale* **2021**, *13*(*46*), 19376−19384.

(46) Brun, Y. Solving NP-complete problems in the tile assembly model, *Theoretical Computer Science* **2008**, *395*(*1*), 31–46.

(47) Kang, H.; Lin, T.; Xu, X. J.; Jia, Q.-S.; Lakerveld, R.; Wei, B. DNA dynamics and computation based on toehold-free strand displacement, *Nature Communications* **2021**, *12*(*1*), 4994.

(48) Yurke, B.; Turberfield, A. J.; Mills, A. P.; Simmel, F. C.; Neumann, J. L. A DNA-fuelled molecular machine made of DNA, *Nature* **2000**, *406*(*6796*), 605–608.

(49) Seelig, G.; Soloveichik, D.; Zhang, D. Y.; Winfree, E. Enzyme-free nucleic acid logic circuits, *Science* **2006**, *314*(*5805*), 1585–1588.

(50) Krishnan, Y.; Simmel, F. C. Nucleic acid based molecular devices, *Angewandte Chemie-International Edition* **2011**, *50*(*14*), 3124–3156.

(51) Shu, J.-J.; Wang, Q.-W.; Yong, K.-Y. Programmable DNA-mediated decision maker, *International Journal of Bio-Inspired Computation* **2017**, *10*(*1*), 51–55.

(52) Simmel, F. C.; Yurke, B.; Singh, H. R. Principles and applications of nucleic acid strand displacement reactions, *Chemical Reviews* **2019**, *119*(*10*), 6326–6369.

(53) Picuri, J. M.; Frezza, B. M.; Ghadiri, M. R. Universal translators for nucleic acid diagnosis, *Journal of the American Chemical Society* **2009**, *131*(*26*), 9368–9377.

(54) de Murieta, I. S.; Rodríguez-Patón, A. DNA biosensors that reason, *BioSystems* **2012**, *109*(*2*), 91–104.

(55) Zhang, C.; Zhao, Y. M.; Xu, X. M.; Xu, R.; Li, H. W.; Teng, X. Y.; Du, Y. Z.; Miao, Y. Y.; Lin, H.-C.; Han, D. Cancer diagnosis with DNA molecular computation, *Nature Nanotechnology* **2020**, *15*(*8*), 709–715.

(56) Schaffter, S. W.; Chen, K.-L.; O'Brien, J.; Noble, M.; Murugan, A.; Schulman, R. Standardized excitable elements for scalable engineering of far-from-equilibrium chemical networks, *Nature Chemistry* **2022**, *14*(*11*), 1224–1232.

(57) Wang, B. Y.; Thachuk, C.; Ellington, A. D.; Winfree, E.; Soloveichik, D. Effective design principles for leakless strand displacement systems, *Proceedings of the National Academy of Sciences of the United States of America* **2018**, *115*(*52*), E12182–E12191.

(58) Chen, X.; Liu, X. Y.; Wang, F.; Li, S. R.; Chen, C. Z.; Qiang, X. L.; Shi, X. L. Massively parallel DNA computing based on domino DNA strand displacement logic gates, *ACS Synthetic Biology* **2022**, *11*(*7*), 2504–2512.

(59) Qian, L. L.; Winfree, E.; Bruck, J. Neural network computation with DNA strand displacement cascades, *Nature* **2011**, *475*(*7356*), 368–372.

(60) Okumura, S.; Gines, G.; Lobato-Dauzier, N.; Baccouche, A.; Deteix, R.; Fujii, T.; Rondelez, Y.; Genot, A. J. Nonlinear decision-making with enzymatic neural networks, *Nature* **2022**, *610*(*7932*), 496–501.

(61) Xie, Z.; Wroblewska, L.; Prochazka, L.; Weiss, R.; Benenson, Y. Multi-input RNAi-based logic circuit for identification of specific cancer cells, *Science* **2011**, *333*(*6047*), 1307–1311.

(62) Morihiro, K.; Ankenbruck, N.; Lukasak, B.; Deiters, A. Small molecule release and activation through DNA computing, *Journal of the American Chemical Society* **2017**, *139*(*39*), 13909–13915.

(63) Benenson, Y.; Gil, B.; Ben-Dor, U.; Adar, R.; Shapiro, E. An autonomous molecular computer for logical control of gene expression, *Nature* **2004**, *429*(*6990*), 423–429.

(64) Benenson, Y. Synthetic biology with RNA: Progress report, *Current Opinion in Chemical Biology* **2012**, *16*(*3-4*), 278–284.

(65) Pardee, K.; Green, A. A.; Takahashi, M. K.; Braff, D.; Lambert, G.; Lee, J. W.; Ferrante, T.; Ma, D.; Donghia, N.; Fan, M.; Daringer, N. M.; Bosch, I.; Dudley, D. M.; O'Connor, D. H.; Gehrke, L.; Collins, J. J. Rapid, low-cost detection of *Zika virus* using programmable biomolecular components, *Cell* **2016**, *165*(*5*), 1255–1266.

(66) Deaton, R.; Garzon, M.; Murphy, R. C.; Rose, J. A.; Franceschetti, D. R.; Stevens, S. E. Reliability and efficiency of a DNA-based computation, *Physical Review Letters* **1998**, *80*(*2*), 417–420.

(67) Tanaka, F.; Kameda, A.; Yamamoto, M.; Ohuchi, A. Design of nucleic acid sequences for DNA computing based on a thermodynamic approach, *Nucleic Acids Research* **2005**, *33*(*3*), 903–911.

(68) Yang, C.-N.; Huang, K.-S.; Yang, C.-B.; Hsu, C.-Y. Error-tolerant minimum finding with DNA computing, *International Journal of Innovative Computing Information and Control* **2009**, *5*(*10A*), 3045–3057.

(69) Saghatelian, A.; Völcker, N. H.; Guckian, K. M.; Lin, V. S.-Y.; Ghadiri, M. R. DNA-based photonic logic gates: AND, NAND, and INHIBIT, *Journal of the American Chemical Society* **2003**, *125*(*2*), 346–347.

(70) Smith, L. M.; Corn, R. M.; Condon, A. E.; Lagally, M. G.; Frutos, A. G.; Liu, Q. H.; Thiel, A. J. A surface-based approach to DNA computation, *Journal of Computational Biology* **1998**, *5*(*2*), 255–267.

(71) Liu, Q. H.; Wang, L. M.; Frutos, A. G.; Condon, A. E.; Corn, R. M.; Smith, L. M. DNA computing on surfaces, *Nature* **2000**, *403*(*6766*), 175–179.

(72) Madhavan, D.; Cuk, K.; Burwinkel, B.; Yang, R. X. Cancer diagnosis and prognosis decoded by blood-based circulating microRNA signatures, *Frontiers in Genetics* **2013**, *4*(*116*).

(73) Green, A. A.; Kim, J. M.; Ma, D.; Silver, P. A.; Collins, J. J.; Yin, P. Complex cellular logic computation using ribocomputing devices, *Nature* **2017**, *548*(*7665*), 117–121.

(74) Choi, S.; Lee, G.; Kim, J. Cellular computational logic using toehold switches, *International Journal of Molecular Sciences* **2022**, *23*(*8*), 4265.

1